\documentclass[prl,twocolumn,showpacs,superscriptaddress,oneside]{revtex4}%
\usepackage{amssymb}
\usepackage{amsfonts}
\usepackage{graphicx}
\usepackage{epsfig}
\usepackage{amsmath}
\begin{document}
\title{Hole and Electron Contributions to the Transport Properties of 
Ba(Fe$_{1-x}$Ru$_{x}$)$_{2}$As$_{2}$ Single Crystals}
\author{F. Rullier-Albenque}
\email{florence.albenque-rullier@cea.fr}
\affiliation{Service de Physique de l'Etat Condens\'e, Orme des Merisiers, IRAMIS, CEA Saclay (CNRS URA 2464), 91191 Gif sur Yvette cedex, France}
\author{D. Colson}
\affiliation{Service de Physique de l'Etat Condens\'e, Orme des Merisiers, IRAMIS, CEA Saclay (CNRS URA 2464), 91191 Gif sur Yvette cedex, France}
\author{A. Forget}
\affiliation{Service de Physique de l'Etat Condens\'e, Orme des Merisiers, IRAMIS, CEA Saclay (CNRS URA 2464), 91191 Gif sur Yvette cedex, France}
\author{P. Thu\'ery}
\affiliation{SIS2M, CEA/CNRS UMR 3299, LCCEf, IRAMIS, 91191 Gif-sur-Yvette cedex, France}
\author{S. Poissonnet}
\affiliation{Service de Recherches de M\'etallurgie Physique, CEA Saclay, 91191 Gif sur Yvette cedex, France}

\date{May 7th 2010}

\begin{abstract}
We report a systematic study of structural and transport properties in single crystals of 
Ba(Fe$_{1-x}$Ru$_{x}$)$_{2}$As$_{2}$ for $x$ ranging from 0 to 0.5. The isovalent substitution 
of Fe by Ru leads to an increase of the $a$ parameter and a decrease of the $c$ parameter, resulting 
in a strong increase of the AsFeAs angle and a decrease of the As height above the Fe planes. Upon Ru substitution, 
the magnetic order is progressively suppressed and superconductivity emerges for $x\geq0.15$, 
with an optimal $T_{c}\simeq20$K at $x=0.35$ and coexistence of magnetism and superconductivity between 
these two Ru contents. Moreover, the Hall coefficient $R_{H}$ which is always negative and decreases 
with temperature in BaFe$_{2}$As$_{2}$, is found to increase here with decreasing $T$ and even change 
sign for $x\geq0.20$. For $x_{Ru}=0.35$, photo-emission studies have shown that the number of 
holes and electrons are similar with $n_{e}=n_{h}\simeq0.11$ carriers/Fe, that is twice larger than found in BaFe$_{2}$As$_{2}$ \cite{Brouet2}. 
Using this estimate, we find that the transport properties of Ba(Fe$_{0.65}$Ru$_{0.35}$)$_{2}$As$_{2}$ 
can be accounted for by the conventional multiband description for a compensated semi-metal. In particular, our results show that the mobility of holes is strongly enhanced upon Ru addition and overcomes that of electrons at low temperature when $x_{Ru}\geq0.15$.

\end{abstract}

\maketitle

\section{Introduction}
In iron pnictides the appearance of high-$T_{c}$ superconductivity induced by carrier doping 
or pressure in close proximity to the
antiferromagnetic (AF) phase, appears very similar to the behavior of cuprates and
heavy-fermion superconductors and has been taken as the signature of unconventional superconductivity 
in these compounds. It seems now well established that magnetism and superconductivity (SC) 
are intimately correlated and directly connected to the peculiar features of the electronic 
structures of these compounds. More precisely the changes in the Fermi surface and the modifications 
of the nesting conditions between the hole and electron pockets have been proposed to be the driving 
force for the suppression of antiferromagnetism and the emergence of superconductivity with 
a sign reversing $s_{\pm}$ symmetry \cite{review_theo}.

So far a lot of studies have been devoted to the 122 family as superconductivity can be induced not 
only by doping with holes \cite{Rotter} or electrons \cite{el_doped, Chu} but also through chemical or 
physical pressure \cite{Ren, Alireza}. Several investigations have been done in order to find out 
a relevant parameter allowing to explain the modifications of the electronic structure 
and the emergence of superconductivity in these different systems. On one hand, it has been 
argued that structural modifications could be more important than doping in achieving superconductivity, either through the height of As with respect to Fe planes \cite{Mizugushi} or the  value of the 
As-Fe-As bonding tetrahedral angle \cite{Kimber}. On the other hand, in the case of electron 
doped compounds, the steric effect due to different atomic substitutions in the Fe planes, 
has been shown to be of minor importance compared to the effect of doping \cite{Canfield}. 
Let us note that the very low level of substitution sufficient to induce SC in this latter case 
is compatible with weak structural distortion effects.
However, in the case of hole doping for which large substitution level (around 35\%) is necessary to get 
the optimal $T_{c}$, it seems more difficult to distinguish between doping and structural modifications. 
    
Studies of transport properties are a priori one of the simplest way to investigate 
the modifications of the electronic structure. However the situation in the 122 family is far from being clear. In the undoped BaFe$_{2}$As$_{2}$ parent, for which the electron and hole contents are identical ($n=n_{e}=n_{h}$), it is found quite surprisingly 
that the Hall coefficient $R_{H}$ is always negative,
indicating that electrons dominate the transport properties both above and below the 
structural/magnetic transition at $T\backsimeq140$K. The same observation has been 
found for Co doped samples 
all over the phase diagram \cite{FRA-PRL, Fang}. For all these compounds, it seems that the holes 
are highly scattered and thus not directly visible in the transport properties \cite{Optics}.
The case of isovalent substitution of As by P is even more intriguing as a negative Hall coefficient 
is also found for all P contents \cite{Kasahara}. In fact, positive $R_{H}$ have been only reported 
for K and Cr doped BaFe$_{2}$As$_{2}$, as naturally expected for hole doped compounds \cite{Hall-BaKFeAs, Sefat-Cr}.
    
The isovalent substitution of Fe by Ru provides another alternative to study the modifications 
of the transport properties in the 122 family. It has been shown recently on 
polycrystalline samples \cite{Paulraj, Schnelle} that the introduction of Ru suppresses the SDW 
magnetic order and induces SC. Density functional calculations show that Ru substitution does 
not induce any charge imbalance between the bands and no additional bands related to Ru appear at the Fermi level \cite{Zhang}. 
However a negative Hall coefficient with a weak T dependence has also been reported for polycrystalline BaFe$_{1.25}$Ru$_{0.75}$As$_{2}$ \cite{Paulraj}.
    
Here we report on structural, resistivity and Hall effect data obtained 
on single crystals of Ba(Fe$_{1-x}$Ru$_{x}$)$_{2}$As$_{2}$ for $x$ ranging from 0 to 0.5. 
As reported previously, we confirm the suppression of the magneto-structural transition and 
the emergence of SC for $x\gtrsim0.15$ with an optimal $T_{c}\sim20K$ at $x\simeq0.35$. We find that 
the lattice parameters $a$ ($c$) respectively increases (decreases) upon Ru addition, so that 
$a/c$ increases markedly. 
As for the transport properties, our Hall effect measurements evidence the 
contribution of both holes and electrons in the transport properties, as the Hall coefficient 
changes sign, from negative to positive with decreasing temperature. On the other hand, Angle 
Resolved Photo-Emission Spectroscopy (ARPES) measurements performed on similar single crystals with 
$x=0.35$ \cite{Brouet2} confirm that Ru-substituted BaFe$_{2}$As$_{2}$ behave 
as a compensated metal with essentially the same number of holes and electrons, i.e. 
$n=n_{e}=n_{h}\simeq0.11$ carriers/Fe. Assuming a two band model to describe the transport properties, 
we show that it is possible here to disentangle the respective contributions of electrons 
and holes. Quite surprisingly the deduced electron and hole resistivity curves 
display similar $T$ dependences as those found respectively for Co-doped and 
K-doped BaFe$_{2}$As$_{2}$ at optimal doping. Their evolutions with Ru content show that the mobility of holes is more affected than that of electrons. The strong modification of 
the electronic structure of BaFe$_{2}$As$_{2}$ with Ru substitution revealed by ARPES \cite{Brouet2} 
might be the key factor for governing these properties.
    
\section{Samples and structural measurements.}

Single crystals of Ba(Fe$_{1-x}$Ru$_{x}$)$_{2}$As$_{2}$ with Ru contents $x$ 
ranging from 0 to 0.50 were grown using a FeAs/Ru+As self-flux method. 
Small Ba chunks, FeAs powder and RuAs (or Ru+As) powders were mixed in the 
ratio Ba:(FeAs+RuAs)=1:4. Starting products were put in an alumina crucible 
and sealed in an evacuated quartz tube which was put into a tubular furnace. The samples 
were heated to 1180$^\circ$C, held at this temperature for 4h, cooled slowly 
first to 1000$^\circ$C (3-6$^\circ$C/h) and then more rapidly to room temperature. 
Clean crystals of typical dimensions 0.5x0.5x0.05 mm$^{3}$ were mechanically 
extracted from the flux. It is worth pointing out here that it is very difficult to get homogeneous 
single crystals for $x_{Ru}\geq0.2$, probably due to the very high melting temperatures of Ru and RuAs with respect to that of FeAs. Consequently, the Ru composition of each studied crystal 
has been determined with a Camebax SX50 electron microprobe in several spots of the surfaces. 
The structural properties were 
characterized by single crystal X-ray diffraction on very thin 
platelets ($\sim0.10$ x 0.05 x 0.01$mm^{3}$). The data were collected at 
room temperature on a Nonius Kappa-CCD area detector diffractometer \cite{Hooft} 
using graphite-monochromated Mo K$\alpha$ radiation ($\lambda = 0.71073$ \AA). 
The data (combinations of $\varphi$- and $\omega$-scans giving complete 
data sets up to $\theta = 27.4$ deg. at least and a minimum redundancy of 4 for 90\% 
of the reflections) were processed with HKL2000 \cite{RX}. 
Absorption effects were corrected empirically with the 
program SCALEPACK \cite{RX}. The structures were refined 
in the tetragonal space group I4/mmm by full-matrix least-squares 
on F2 with SHELXL-97 \cite{Sheldrick}. 
All the atoms were refined with anisotropic displacement parameters, 
so that 10 parameters were refined on $\sim100$ independent reflections. 
Fe and Ru were constrained to retain the same displacement parameters, 
which enabled to refine the Ru content $x$. 
The final $R_{1}$ indices are in the range $0.020-0.04$2 and the $wR_{2}$ values 
in the range $0.043-0.113$ \cite{formulaRX}.

\begin{figure}
\centering
\includegraphics[width=8cm]{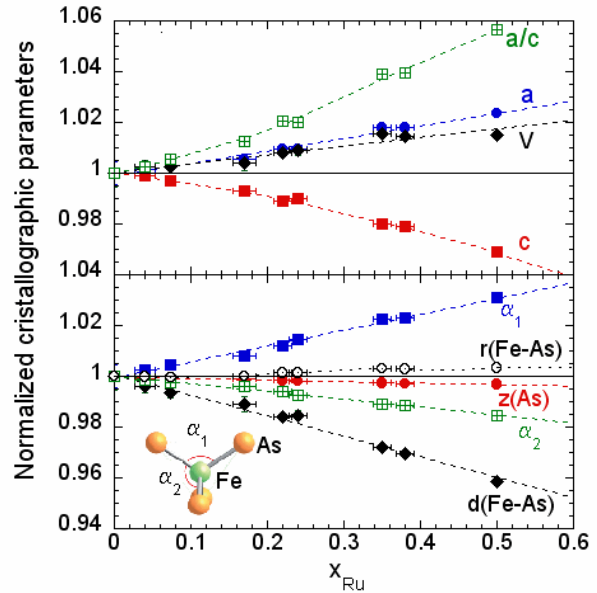}
\caption{(color on line) (a)Variation of the lattice parameters $a$ and $c$, their ratio $a/c$ and the 
unit cell volume V, normalized 
to their values for the undoped compound, as a function 
of the Ru content $x$ (determined by X-rays diffraction). (b) The same for different 
crystallographic parameters: the two AsFeAs tetrahedral angles, $\alpha_{1}$ and $\alpha_{2}$, the 
c-axis coordinate of As, $z_{As}$, the Fe-As interatomic distance $r(Fe-As)$ and the height of As above Fe layers, $d_{Fe-As}$.}
\label{Fig.1}
\end{figure}

\begin{table}
\caption{Crystallographic data taken at room temperature for the parent and 
Ru-substituted (x= 0.38) Ba(Fe$_{1–x}$Ru$_{x}$)$_{2}$As$_{2}$. The space group of both compounds is I4/mmm and 
the atomic coordinates are: Ba(0,0,0), Fe/Ru(0.5,0,0.25) and As (0,0,z). $\alpha_{1}$ 
and $\alpha_{2}$ are the bonding tetrahedral angles as sketched in Fig.1.}
\label{tab:table1}
\begin{ruledtabular}
\begin{tabular}{ccc}
 &\textbf{x=0}&\textbf{x=0.38}\\\hline
a (\AA) &3.9633(4) &4.0342(5)\\
c (\AA) &13.022(2) &12.749(2)\\
V (\AA$^{3}$) &204.55(4) &207.49(5)\\
$z_{As}$ &0.35424(6) &0.35328(8)\\
$\alpha_{1}$ x 2 (deg) &111.18(3) &113.73(4)\\
$\alpha_{2}$ x 4 (deg) &108.624(15) &107.39(2)\\
As height (\AA) &1.3575 &1.3165\\
Fe-As interatomic distance (\AA) &2.402(2) &2.409(2)\\
FeAs layer spacing (\AA) &3.796 &3.741\\
\end{tabular}
\end{ruledtabular}
\end{table}

The relative lattice parameters are plotted in Fig.1 
versus the refined value of $x_{Ru}$, while numerical data are reported 
in table 1 for $x=0$ (undoped BaFe$_{2}$As$_{2}$) and $x=0.38$ (optimal doping). 
Upon substitution, the $a$ parameter and the cell volume increase while the $c$ parameter 
decreases in about the same proportion (by ~2-3\% for $x=0.5$) in agreement 
with data on polycrystalline samples \cite{Paulraj}. The main effect of Ru 
substitution is thus 
to strongly increase the ratio $(a/c)$. This results in a strong increase of the As-Fe-As tetrahedral 
bonding angle $\alpha_{1}$ displayed in fig.1(b) which varies by 4\% for $x=0.5$. Moreover, 
while the $z$ parameter of As and the Fe-As interatomic distance are nearly unaffected by Ru substitution, 
the vertical distance of As from the Fe layers decreases, due to the strong $c$ decrease. 
These tendencies can be explained quite naturally by the larger size of 
Ru$^{2+}$ compared to Fe$^{2+}$ which expands the distances within Fe-Ru planes. Also the larger delocalization of the Ru 4d orbitals with respect to Fe 3d reinforces the hybridization with As and 
then reduces the $c$ axis parameter \cite{Zhang}. 
Let us note that the variation of the lattice parameters upon Ru substitution displays the same trend 
as that observed in electron doped BaFe$_{2}$As$_{2}$ compounds, although the incidence on the structure 
appears much weaker in this latter case \cite{Canfield}. However it is at odds to the 
tendency found for hole doping 
or pressure \cite{Rotter_2,Kimber} for which both $a$ and the As-Fe-As 
$\alpha_{1}$ angle are found to decrease. 

\section{Resistivity measurements and phase diagram}

Transport measurements were performed on crystals cleaved to thicknesses 
lower than 20 $\mu$m and cut to get square samples with $\sim0.2-0.3$ mm width (See fig.2(c)) 
Contacts were done with silver epoxy in the Van der Pauw configuration \cite{VdP}.

\begin{figure}
\centering
\includegraphics[width=8cm]{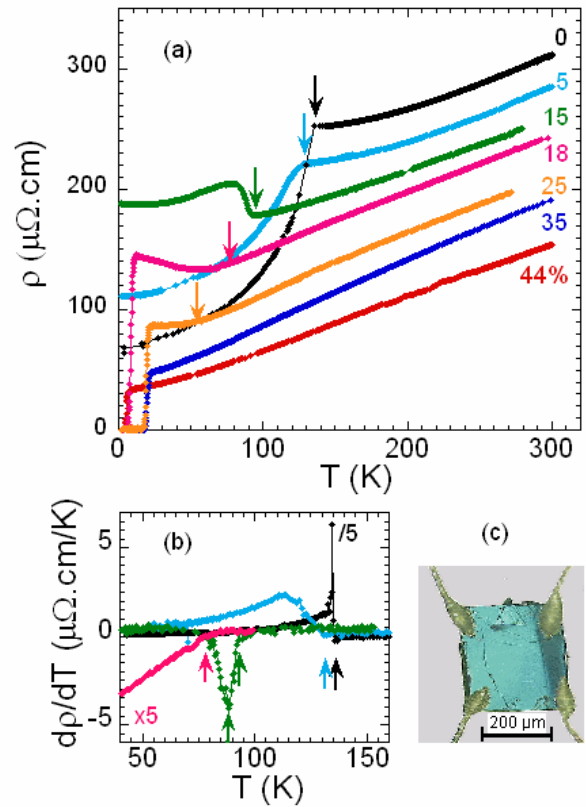}
\caption{(color on line) (a) Temperature dependence of the in-plane resistivity $\rho (T)$ 
of Ba(Fe$_{1-x}$Ru$_{x}$)$_{2}$As$_{2}$ single crystals. The arrows point the strong anomalies in the $\rho (T)$ curves that signal the occurrence of the magneto-structural transition. They are determined more precisely in (b) which shows $d\rho/dT$ versus $T$. For the $x=0.26$ sample, the arrow corresponds to the maximum of the Hall coefficient as explained in the text. (c) Picture of a typical sample mounted in the Van der Pauw configuration.}
\end{figure}

Fig.2(a) shows the $\rho(T)$ curves as a function of Ru content. 
In BaFe$_{2}$As$_{2}$, the combined structural and magnetic (S-M) transition 
at T$_{0}$=137 K is signalled by a decrease of the resistivity \cite{anomaly-T_SDW}. 
For the lowest Ru content $x=0.05$ studied here, we still observe a resistivity 
decrease at the S-M transition but with a much wider transition as seen in fig.2(b) 
which shows the resistivity derivative $d\rho/dT$. Then for larger Ru contents, 
the resistive signature of the S-M transition, determined by the deviations in $d\rho/dT$, changes shape towards a step-like increase 
of the resistivity as observed in electron doped compounds. 
In contrast, let us point out that the S-M transition is always signalled 
by a decrease of the resistivity in polycrystalline 
samples \cite{Paulraj, Schnelle}, which might be related to inhomogeneities in the samples or directional averaging due to different in-plane and out of plane resistivity variations at the S-M transition. 
For $x=0.15$, it is possible to distinguish between the structural and magnetic
transitions, using the same criteria based on the variation of $d\rho/dT$ as proposed 
for Co-doped samples (arrows in fig.2(b)) \cite{Chu, Pratt}. This would give 95 and 88K respectively 
for $T_{S}$ and $T_{SDW}$. For the other samples, this distinction is not possible and microscopic investigations are needed to determine the respective values of $T_{S}$ and $T_{SDW}$.

Superconductivity appears for $x\geq 0.15$ and a
maximum $T_{c}$ of $19.5\pm 0.5$K is found for $x=0.35$, in excellent agreement with the value found for polycrystalline samples of Ru substituted Ba(Sr)Fe$_{2}$As$_{2}$ \cite{Paulraj, Schnelle}. 
The $T-x$ phase diagram obtained for Ba(Fe$_{1-x}$Ru$_{x}$)$_{2}$As$_{2}$ 
is displayed in fig.3. One can notice that the superconducting region is rather small, 
particularly in the "overdoped" region where $T_{c}$ is found to
drop rapidly beyond $x=0.35$. As in the doped compounds, coexistence between magnetism
and superconductivity is clearly evidenced, with the optimal $T_{c}$ occurring when long range magnetic order is fully suppressed. This points here again to the importance of spin fluctuations for superconductivity in these systems. From these resistivity data, it is not possible to know whether the  
coexistence between magnetism and superconductivity occurs at the atomic scale like in Co-doped samples \cite{Laplace} or comes from 
phase segregation as observed in K-doped samples \cite{K-muSr,K-RMN} and further studies at the miscoscopic level are needed to assess this point.

\begin{figure}
\centering
\includegraphics[width=8cm]{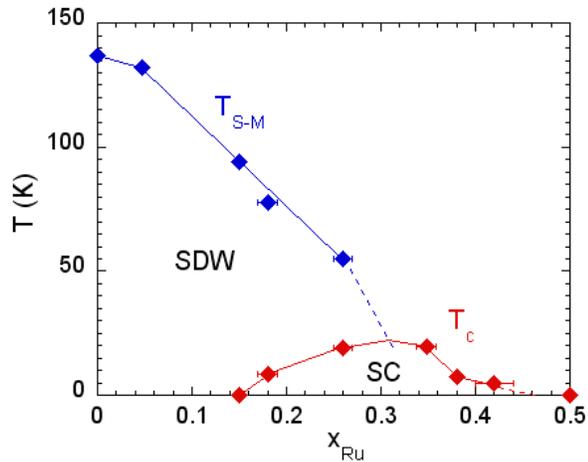}
\caption{(color on line) Phase diagram of the Ba(Fe$_{1-x}$Ru$_{x}$)$_{2}$As$_{2}$ system showing the evolution of the structural-magnetic transition $T_{S-M}$ and critical temperatures $T_{c}$ versus Ru content $x_{Ru}$. Values of $T_{c}$ are taken at the mid-point of superconducting transitions. The sample with $x=0.5$ is no longer superconducting as shown by SQUID magnetometer measurements.}
\label{Fig.3}
\end{figure}

We find here that superconductivity is induced upon Ru addition while the FeAs$_{4}$ tetrahedra become strongly distorted as both the $\alpha_{1}$ and $\alpha_{2}$ angles deviate from 109.5 deg, the ideal tetrahedral value. This observation conflicts with the claim that the regularisation of tetrahedra is the optimal condition for achieving superconductivity in pnictides \cite{Lee, Zhao, Kimber}. Let us also note that Ru substitution in the 1111 PrFeAsO compound induces similar crystallographic modifications as those observed here in BaFe$_{2}$As$_{2}$ \cite{McGuire} with suppression of the magnetic order but no apparition of superconductivity.

On another hand, band structure calculations have pointed out the important impact of the vertical distance $d_{Fe-As}$ on the Fermi surface topology of iron pnictices \cite{Vildosola}. 
Mizugushi \textit{et al.} \cite{Mizugushi} have recently shown that a striking correlation 
between $T_{c}$ and the Fe-As distance is followed by a lot of different FeAs superconductors. 
This plot is symmetric with a peak around $d_{Fe-As}$=1.38\AA. We find that the point corresponding 
to Ba(Fe$_{1.62}$Ru$_{0.35}$)$_{2}$As$_{2}$ ($d_{Fe-As}$=1.3165\AA) and $T_{c}\simeq20$K is on 
the left branch while the one for pressure or hole doping is located on the right one. Even though other factors are clearly at play for governing the apparition of superconductivity in the 122 family, the relationship between the values of $T_{c}$ and d$_{Fe-As}$ may provide a helpful hint to 
understand the modifications of the electronic properties.

\section{Hall effect and analysis of transport properties}

The temperature dependences of the Hall coefficient $R_{H}$  
are displayed in Fig.4 for different Ru concentrations. In the paramagnetic state of the samples, 
we have checked that the Hall resistivity is always linear in field up to 14T, which allows to define 
$R_{H}$ unambiguously \cite{mag_state}. This linearity is illustrated  in the inset of 
Fig.4  for the $x=0.35$ sample. The strong reduction in the Hall coefficient 
at the S-M transition is well correlated to the anomalies seen in $d\rho/dT$ and represented by 
arrows in the figure. It can be associated, as in the undoped parent, to the reduction of 
carrier density due to the reconstruction and/or partial gaping of the Fermi surfaces. The fact that $R_{H}$ remains negative indicates that electrons still dominate the transport properties 
in the magnetic phase of Ru-substituted samples.

\begin{figure}
\centering
\includegraphics[width=8cm]{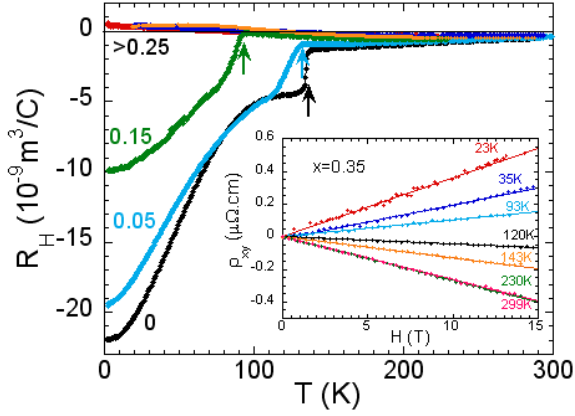}
\caption{(color on line) $T$ dependence of the Hall coefficient $R_{H}(T)$
for various compositions. The temperatures at which $R_{H}$ starts decreasing due to the apparition 
of the M-S transition correspond exactly to those where anomalies are seen in the 
resistivity curves(arrows). The inset shows that the Hall resistivity $\rho_{xy}$ of the $x=0.35$ 
sample is linear in magnetic field up to 14T whatever $T$. A sign change of the slope occurs 
between 93 and 120K.}
\label{Fig.4}
\end{figure}

\begin{figure}
\centering
\includegraphics[width=8cm]{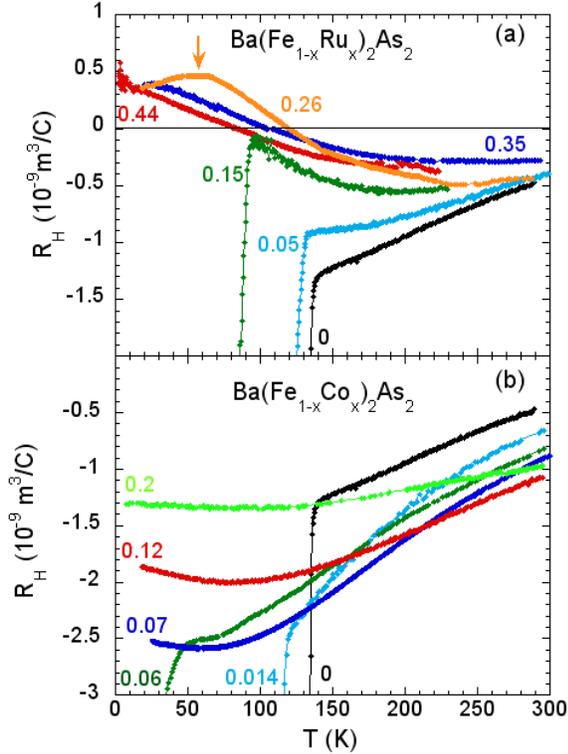}
\caption{(color on line) $T$ dependence of the Hall coefficient $R_{H}(T)$
in the vicinity of $R_{H}=0$ for (a) Ba(Fe$_{1-x}$Ru$_{x}$)$_{2}$As$_{2}$ and (b)
Ba(Fe$_{1-x}$Co$_{x}$)$_{2}$As$_{2}$, from ref.\cite{FRA-PRL}.} 
\label{Fig.5}
\end{figure}

Fig.5(a) shows an enlarged view of the evolution of the Hall coefficient 
in the paramagnetic phase. For comparison, we have also plotted in Fig.5(b) similar data obtained for Ba(Fe$_{1-x}$Co$_{x}$)$_{2}$As$_{2}$ at various dopings \cite{FRA-PRL}. In this latter case, $R_{H}$ is always found negative, indicating that the contribution of electrons dominate the transport properties. 
However an opposite trend appears 
as soon as Ru is added to BaFe$_{2}$As$_{2}$. For $x=0.15$, $R_{H}$ nearly reaches zero 
before dropping at the S-M transition 
and for higher Ru contents, a change of sign of $R_{H}$ occurs at low temperature. In particular for the  $x\sim0.25$ sample, we observe that $R_{H}$ increases and becomes positive upon 
cooling and then appears to slightly decrease again for $T\leq50K$. This can be related to 
the flattening of the $\rho(T)$ curves observed in the same temperature range and this is for 
us the sign that the S-M transition takes place at $T\simeq50$K in this sample.

In multiband systems, it is well known that a temperature variation of the Hall coefficient can 
be assigned to different variations of hole and electron mobilities with temperature. The observation 
of a sign change of $R_{H}$ in Ba(Fe$_{1-x}$Ru$_{x}$)$_{2}$As$_{2}$ indicates that holes and 
electrons contribute similarly to the transport in a large temperature range. More precisely ARPES data 
on crystals with 35\% Ru \cite{Brouet2} have shown
that the number of holes and electrons are similar, i.e. $n=n_{e}=n_{h}\simeq0.11$ carriers/Fe. It is worth pointing out that this value is significantly larger 
than that determined by ARPES  in the paramagnetic phase of BaFe$_{2}$As$_{2}$: $n=0.06(2)$ carriers/Fe \cite{Brouet1}, which indicates that even though Ru is isovalent of Fe, it 
induces important modifications of the electronic structure. This equality of $n_{e}$ and $n_{h}$ is consistent with the 
observation that the Hall resistivity $\rho_{xy}$ is always linear with magnetic field. Indeed in a two band model, the Hall resistivity $\rho_{xy}$ can be written out as:
\begin{equation}
\rho_{xy}=\frac{1}{e}\dfrac{n_{h}\mu_{h}^{2}-n_{e}\mu_{e}^{2}+(\mu_{h}\mu_{e})^{2}(n_{h}-n_{e})H^{2}}{(n_{h}\mu_{h}+n_{e}\mu_{e})^{2}+(\mu_{h}\mu_{e})^{2}(n_{h}-n_{e})^{2}H^{2}}H \label{1}
\end{equation}
where $\mu_{h}=|e|\tau_{h}/m_{h}$ ($\mu_{e}=|e|\tau_{e}/m_{e}$) are the mobilities of 
holes (electrons) and $\tau_{h}$ ($\tau_{e}$) and $m_{h}$ ($m_{e}$) their relaxation rates and 
effective masses.
For $n_{e}=n_{h}=n$, the $H^{2}$ term in the numerator and denominator of Eq.(1) vanishes resulting in linear variation of $\rho_{xy}$ with $H$, whatever $H$ and $T$. 

Knowing $n$, it is then straightforward to deduce the respective contributions 
of electrons of holes to the transport for the $x=0.35$ sample from the resistivity and Hall coefficient 
data, using:
\begin{equation}
1/\rho=\sigma=\sigma _{e}+\sigma _{h}  \label{2}
\end{equation}
and
\begin{equation}
R_{H}=\frac{1}{ne}\frac{\mu _{e}-\mu _{h}}{\mu _{e}+\mu _{h}}=\frac{1}{ne}\frac{\sigma _{e}-\sigma _{h}}{\sigma _{e}+\sigma _{h}}  \label{3}
\end{equation}
The resulting resistivity curves obtained for electrons and holes are displayed in Fig.6(a). 
It is striking to see that the shapes of the curves resemble those obtained 
respectively for electron and hole doped BaFe$_{2}$As$_{2}$ at optimal doping. These similarities give strong support to the validity of the decomposition using the two band model. One can also notice that
$\rho _{h}(T)$ displays a nearly $T^2$ dependence as in Ba$_{0.65}$K$_{0.35}$Fe$_{2}
$As$_{2}$ \cite{Hall-BaKFeAs} while $\rho _{e}(T)$ exhibits a nearly
linear T-dependence up to 150K as in Ba(Fe$_{0.93}$Co$_{0.07}$)$_{2}$As$_{2}$
\cite{FRA-PRL}. These temperature variations appear then tightly connected to the type of carriers - hole or electrons - and indicate an intrinsic disparity between their respective properties. 

It has been suggested that the linear T dependence of resistivity found 
in Ba(Fe$_{1-x}$Co$_{x}$)$_{2}$As$_{2}$ or BaFe$_{2}$As$_{1-x}$P$_{x}$ near optimal doping might be 
a general property of unconventional superconductors near a SDW instability \cite{Doiron} or a signature 
of non Fermi liquid behavior \cite{Kasahara}. The observation of a linear behavior for the electrons 
and \textit{not for the holes} in the same sample clearly addresses the question of the real physical origin 
of this linearity. In Ba(Fe$_{1-x}$Co$_{x}$)$_{2}$As$_{2}$, the analysis of combined data of resistivity 
and Hall effect for $n_{e}>n_{h}$ led us to suggest that this linearity comes from an artefact due to 
a small variation of $n_{e}$ with temperature and that the scattering rates obey the $T^{2}$ behavior 
expected for Fermi liquids \cite{FRA-PRL}. 

\begin{figure}
\centering
\includegraphics[width=8cm]{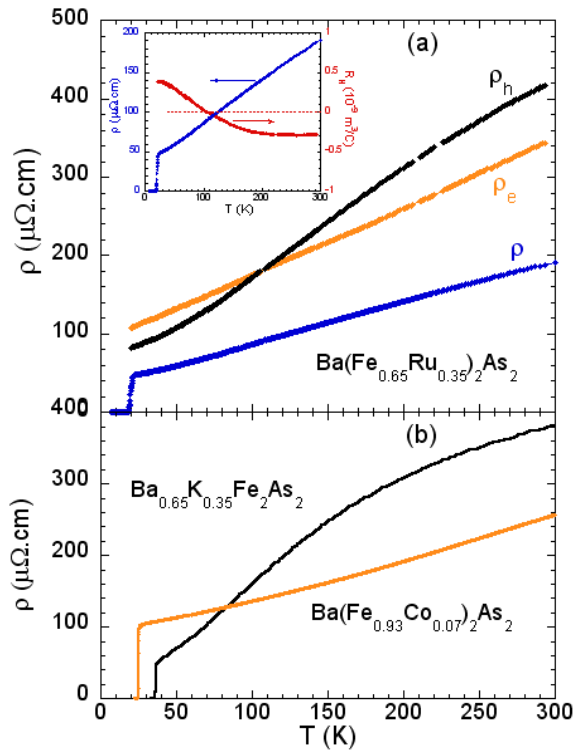}
\caption{(color on line) (a) Respective resistivities of electrons and holes for
Ba(Fe$_{0.65}$Ru$_{0.35}$)$_{2}$As$_{2}$ obtained from the data of
resistivity and Hall coefficient using Equations (2) and (3) with $n_{e}=n_{h}=0.11$, as given by 
ARPES measurements on similar samples \cite{Brouet2}. The raw data for resistivity and Hall coefficient 
are recalled in the inset. (b) The 
resistivity curves for Ba(Fe$_{0.93}$Co$_{0.07}$)$_{2}$As$_{2}$ \cite{FRA-PRL} and Ba$_{0.65}$K$_{0.35}$Fe$_{2}$As$_{2}$ \cite{Hall-BaKFeAs} 
are given for comparison.}
\label{Fig.6}
\end{figure}

\section{Contribution of electrons and holes to the electronic transport}

The most striking result of this study is that electrons and holes contribute similarly to the transport in Ba(Fe$_{1-x}$Ru$_{x}$)$_{2}$As$_{2}$. This is in contrast to the observations of negative Hall coefficient in the undoped parent \cite{FRA-PRL, Fang} or upon isovalent exchange of As by P \cite{Kasahara}. An important question is thus to understand why the holes are more scattered than the electrons in these latter cases.
 
In BaFe$_{2}$As$_{2}$, ARPES data have shown that the number of carriers is $n_{e}=n_{h}=0.06(2)$ carriers/Fe, about twice smaller than the estimate given by LDA calculations 
\cite{Brouet1, Ma}. We have then performed the same decomposition  for BaFe$_{2}$As$_{2}$ in 
the paramagnetic phase, i.e. for $T>140$K, as done above, using either $n = 0.06(2)$ carriers/Fe or $n = 0.15$ carriers/Fe (Fig.7).
It is clear that the results are much more sensitive to the actual value 
of $n$ for the holes than for the electrons: the electron resistivity always displays a metallic 
behavior with similar values while the hole one tends towards a semiconducting behavior when approaching the magneto-structural transition. The disparity between the two types of carriers appears more or less pronounced depending on the value taken for the carrier number.

\begin{figure}
\centering
\includegraphics[width=8cm]{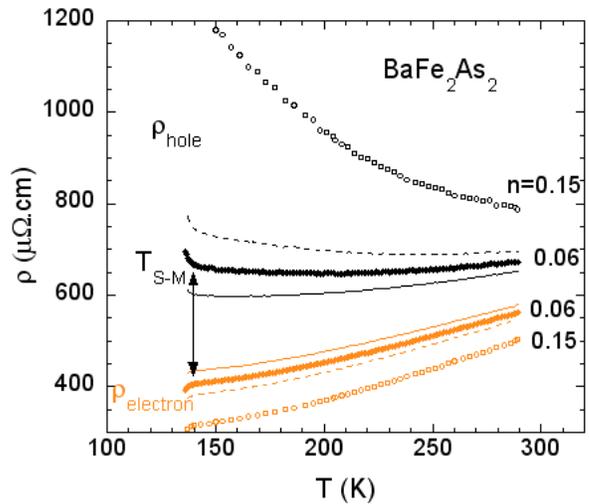}
\caption{(color on line) Respective resistivities of electrons and holes for
BaFe$_{2}$As$_{2}$ in the paramagnetic phase ($T>T_{SM}$) obtained from
resistivity and Hall coefficient data using Equations (2) and (3) using different estimates for the carrier number. The full symbols are for $n=n_{e}=n_{h}=0.06$ carriers/Fe as determined by ARPES measurements \cite{Brouet1} while the full and dotted lines correspond to the extremal values of $n$ due to the $\pm0.02$ error bar. The empty symbols are for the LDA estimate $n=n_{e}=n_{h}=0.15$ carriers/Fe.}
\label{Fig.7}
\end{figure}

It has been suggested that spin fluctuations due to interband electron-hole scattering might play a crucial role to explain the asymmetric behaviors of holes and electrons in undoped and electron doped BaFe$_{2}$As$_{2}$ \cite{Fang}. On the contrary, our Hall coefficient data displayed in Fig.5(a) for different Ru contents seem to indicate that the proximity of magnetism does not play here an important role on the respective mobilities of the carriers. This is more visible in Fig.8 where the decompositions in $\rho_{e}$ and $\rho_{h}$ are reported for different Ru contents, assuming a linear variation of $n$ with $x_{Ru}$. 

\begin{figure}
\centering
\includegraphics[width=8cm]{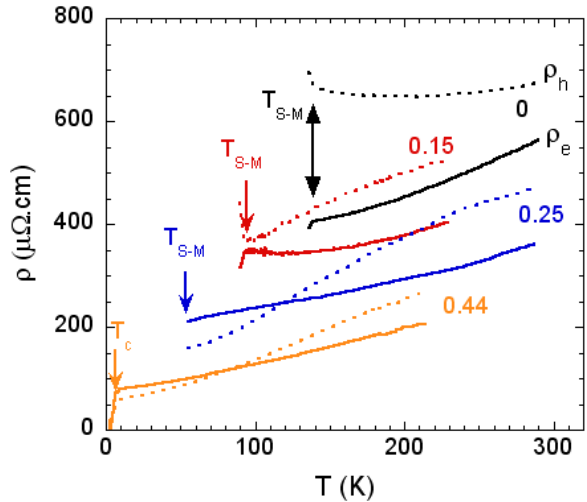}
\caption{(color on line) Same decompositions as that performed in fig.6(a) for different Ru contents, assuming a linear variation of the number of carriers with $x$: $n=0.06+0.14x$ carriers/Fe. Full and dotted lines are for $\rho_{e}$ and $\rho_{h}$ respectively. For the sake of clarity, the data corresponding to $x=0.35$ (fig.6(a)) are not reported in this plot}
\label{Fig.8}
\end{figure}

Even though this analysis is tentative, it gives some trends on the evolution of the transport properties 
of BaFe$_{2}$As$_{2}$ upon Ru addition. In particular, one can notice that the decrease of the hole 
and electron resistivities cannot be entirely explained by the increase of $n$. This therefore points to 
a concomitant increase of their respective mobilities. As shown by ARPES measurements on the $x=0.35$ sample, Ru substitution strongly modifies the electronic structure with respect to that of undoped BaFe$_{2}$As$_{2}$: not only the number of carriers 
has doubled but also the Fermi velocities have increased by a factor 2 or 3 \cite{Brouet2}. 
In fact the electronic structure of Ba(Fe$_{0.65}$Ru$_{0.35}$)$_{2}$As$_{2}$ can be reasonably accounted 
for by LDA calculations with negligible electron correlation effects. This tendency is not observed for electron or hole doped compounds with similar $T_{c}$ \cite{Brouet1, Ding, Yi} and  seems then to be a specific feature of this Ru substituted system. One might reasonably think that these modifications would be at the origin of the evolution of the transport properties observed here. 

Nevertheless the way how these band structure modifications can also affect the strength of spin fluctuations or the carrier mobilities is not clear at present. Moreover, the resemblance displayed in fig.6 between the resistivity curves found for electrons and holes and those measured in electron and hole doped compounds appears very puzzling, as it suggests that the transport properties of electrons and holes are defined by their own and are not tightly dependent on the system of interest. 

\section{Conclusion}
The results presented here and in ref.\cite{Brouet2} clearly show that Ru is isovalent of Fe. 
We have confirmed that Ru substitution suppresses the magnetic state and 
induces superconductivity, which coexist in a given concentration range. 
Therefore it appears qualitatively very similar to the other types of substitutions.
 
From the structural point of view, we have shown that superconductivity can be induced although 
the FeAs$_{4}$ tetrahedra are strongly distorted upon Ru addition. These structural modifications are in total contrast to those induced under pressure or by hole doping. This thus demonstrates that the 
regularisation of tetrahedra cannot be the key structural factor for the occurrence of superconductivity 
as proposed recently \cite{Kimber}. However, we find that the relationship between the optimal $T_{c}$ and the anion height above the Fe planes obeys the same plot as found for a lot of different iron based compounds \cite{Mizugushi}. Even though this cannot be considered as the only parameter to drive superconductivity, this indicates that subtle details of crystal structure might tune specific properties of the Fermi surface necessary to optimize it. 

We have demonstrated that a two band model approach is a prerequisite to get insight into the respective contribution of electrons and holes to the transport properties of these multi-band materials. Using combined studies of transport and ARPES measurements on the same samples, we have been able to 
disentangle the respective contributions of electrons and holes to transport properties. We have evidenced 
that their mobilities become comparable upon Ru addition, even in the close proximity 
to magnetism. In addition, we find that the mobility of holes overcomes that of electrons at low $T$ in superconducting samples. The observation that the $\rho(T)$ curves deduced for electrons and holes are very 
similar to those measured in electron or hole doped compounds suggests that the occurrence of 
optimal $T_{c}$ in all these compounds is linked with well defined features of the electron and hole bands. 
Further work, specifically studying the strength and the evolution of antiferromagnetic spin 
fluctuations with Ru contents, will hopefully allow to clarify the incidence of spin fluctuations, 
electronic correlations and filling of the electronic bands on the 
transport properties of these compounds.

\begin{acknowledgments}
We would like to acknowledge H. Alloul and V. Brouet for fruitful discussions and critical reading of the manuscript. 
\end{acknowledgments}


\begin{thebibliography}{99}
\bibitem{mag_state} Below $T_{S-M}$, the Hall resistivity is no longer linear with magnetic field. The Hall coefficient is then defined as: $R_{H}=\rho_{xy}(14T)/14$.
\end{thebibliography}
\end{document}